# Towards a Natural Language Query Processing System

Chantal Montgomery
School of Computing
Queen's University
Kingston, Canada
15clm1@queensu.ca

Haruna Isah
School of Computing
Queen's University
Kingston, Canada
h.isah@cs.queensu.ca

Farhana Zulkernine
School of Computing
Queen's University
Kingston, Canada
farhana@cs.queensu.ca

*Abstract* — Tackling the information retrieval gap between non-technical database end-users and those with the knowledge of formal query languages has been an interesting area of data management and analytics research. The use of natural language interfaces to query information from databases offers the opportunity to bridge the communication challenges between end-users and systems that use formal query languages. Previous research efforts mainly focused on developing structured query interfaces to relational databases. However, the evolution of unstructured big data such as text, images, and video has exposed the limitations of traditional structured query interfaces. While the existing web search tools prove the popularity and usability of natural language query, they return complete documents and web pages instead of focused query responses and are not applicable to database systems. This paper reports our study on the design and development of a natural language query interface to a backend relational database. The novelty in the study lies in defining a graph database as a middle layer to store necessary metadata needed to transform a natural language query into structured query language that can be executed on backend databases. We implemented and evaluated our approach using a restaurant dataset. The translation results for some sample queries yielded a 90% accuracy rate.

## I. INTRODUCTION

In the real world, humans communicate using natural languages such as English or French. The query-response cycle in a human-to-human communication is often very effective since the person responding to the query can ask for further clarification if the query is not clear. This is, however, different in human-to-computer settings such as querying a database [1]. While databases have been around for decades, query languages for accessing such databases are unlikely to ever become common knowledge for the average end-user. For instance, Structured Query Language (SQL), despite its expressiveness, may hinder users with little or no relational database knowledge from exploring and making use of the data stored in an RDBMS [2]. Furthermore, different databases have different query languages and require that the user understand the exact schema of the database and the roles of various entities in a query [3, 4]. These challenges have led to an increasing interest in research and development of tools such as the natural language interface to databases to enhance human-to-database communications.

Commonly used natural language query interfaces according to Li and Jagadish [3] include (i) keyword-based search interfaces such as Google Scholar, a web-based search engine that indexes the full text or metadata of scholarly literature across many publishing formats and disciplines, (ii) form-based interface such as Web of Science and Scopus in which users first select fields such as topic or author and then type appropriate values for each field, and (iii) visual query builder, a web-based framework that helps researchers in various domains search through database records to identify and correlate data based on semantic concepts. Besides, keywords are insufficient in conveying complex query intent, form-based interfaces are only suitable in cases where queries are predictable and limited to the encoded logic, while visual query builders require users to have extensive knowledge of the schema [3].

A Natural Language Interface to Databases (NLIDB) is a system that allows users to access information stored in a database by typing requests expressed in some natural language such as English [6, 7]. NLIDBs are designed to simplify the interaction between users and computers. Through a natural language interface, users can express queries using natural language and get relevant results in one step without the need to fill out forms or trying different keywords which only returns a ranked list of relevant documents instead of a concise reply containing the specific information [5]. NLIDB enables the retrieval of useful information from any database without the knowledge of specific query languages such as Structured Query Language (SQL) for relational databases [8].

Query-response task in NLIDB is often approached by mapping natural language queries to logical forms or programs that provide the desired response when executed on a database [4]. These interfaces use intermediate representation languages to parse and transform the query from users to formal languages supported by the database [9]. Modern NLIDB systems are increasingly leveraging recent advances in deep learning to parse and translate natural language queries to a corresponding query language such as SQL query over a given database [10]. A major limitation, however, is that training data is assumed to have been acquired a priori and crafted to be well-representative of the types of queries one might ask in the target domain [2].

NLIDBs rely on techniques such as pattern matching, syntactic parsing, and semantic grammar interpretation for natural language queries [3, 4]. Research and development efforts in NLIDB were initially focused on relational databases which are useful in storing structured information, however, there is currently an increasing interest in building natural language interfaces for non-relational databases such as RDF-triple stores or knowledge bases and graph databases [5]. Other existing studies surveyed by Affolter et al. [10] focus on generating a distribution of data values stored in the databases to match values in the user queries to database field names to construct SQL queries. However, as we are currently in the era of big data, such approaches of generating a subset of possible values by applying statistical distribution methods have become impractical and inefficient. This study, therefore, focuses on the development of an interface to a backend relational database for translating natural language queries to SQL queries. Related information about the relational database is kept in a graph database to extend the backend to support multiple distributed databases in the future and be able to compose a query that joins fields from multiple data sources.

Query over graph databases is increasingly attracting much attention [13]. Storing and managing connected semi-structured datasets within relational databases is very challenging because relational databases were originally designed to store and process data in tabular structures. The strength of relational databases lies in their abstraction, however, in practice, maintaining foreign key constraints and computing many JOINs becomes prohibitively expensive [12]. The underlying data layout in graph databases usually does not follow the fixed schema based on tables that implement relations. Multiple types of relational and complex data can be mapped and organized in a non-rigid structure in graph databases [14]. The benefit of using a graph database is the ability to quickly traverse nodes and relationships to find relevant data [11].

*A. Use Case Scenario*

The use case considered in this study is a large restaurant with an existing relational database which is the primary system for storing all transaction records. The data is linked in nature and the restaurant is looking to optimize its data management strategies by developing a cloud-based interface for its customers to effectively access and query information from its database. The proposed interface is aimed at answering questions such as: which restaurants have excellent ratings?

This natural language query should be translated into the following SQL query:

SELECT DISTINCT restaurant_name FROM restaurants WHERE rating_text= "excellent";

An example of the expected result should look as follows:

| Restaurant_name | City | Average_rating |
|---|---|---|
| Atlantic Dishes | Kingston | 4.8 |
| Northern Buffet | Ottawa | 4.7 |
| Lunch Basics | Toronto | 4.6 |

*B. Key Contributions*

The key contributions in this study include (i) background concepts on NLIDB design strategies, (ii) literature review on translating natural language inputs into SQL queries, and (iii) design and implementation of a 3-layered architecture for executing natural language queries on a relational database. The first layer is a cloud-based text entry platform for the users to enter the query text. The middle layer consists of a graph database and algorithms to transform the natural language query into an SQL query. Finally, the third layer consists of a relational database to run the SLQ query on. The novelty of this study is the use of a graph database to store the schema of the backend databases in a way to enable graph search for semantic matching of the natural language query text with database field names. Additional algorithms use the search results and predefined SQL query templates to transform the user query into a SQL query.

*C. Organization*

The paper is organized as follows. Section II presents a background study on approaches to designing NLIDBs and a literature review of recent studies on natural language interfaces to graph databases. Section III presents the architecture and describes the components of the cloud-based interface. Section IV provides details about the implementation and evaluation of the proposed system. Finally, Section V presents concluding remarks and a list of future work.

## II. BACKGROUND STUDY

*A. NLIDB Research Challenges*

There have been numerous attempts towards supporting arbitrary natural language query processing on databases [15]. The use of natural language interfaces for querying databases offers the opportunity to bridge the technological gap between end-users and systems that use formal query languages [6]. The key research problems in this area are depicted in Fig. 1.

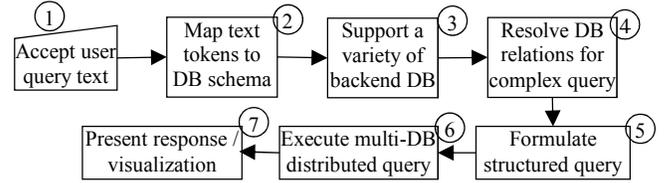

Fig. 1. NLIDB research challenges.

User queries accepted through voice or chat must be (1) first transformed into natural language text from which (2) word tokens have to be extracted and mapped to the backend database schema. Typically, organizations have a hybrid distributed storage system and (3) ideally queries should support existing storage architecture. (4) The next challenge is to find DB relations to join DBs through one or more subqueries and deduce the response. (5) Once relations are mapped, queries must be formulated using appropriate languages for specific storage systems and (6) executed in a distributed manner to optimize query response. (7) Finally, responses must be presented using the preferred format and visualization tools.

*B. Literature Survey*

Affolter et al. [10] identified five different approaches to designing NLIDBs: (i) Keyword-based, (ii) Pattern-based, (iii) Parsing-based, (iv) Grammar-based, and (v) Neural machine translation-based approaches as described below.

*1) Keyword-based*

The keyword-based approach is the most widely used interface for information retrieval [5]. At the core of the keyword-based NLIDB is a token lookup step where the system tries to match the given keywords against an inverted index of the base and metadata [13].

*2) Pattern-based*

The pattern-based NLIDB is an extension of the keyword-based approach with natural language patterns for answering more complex questions such as concepts or aggregations. This approach focuses on the optimization of user interaction.

*3) Parsing-based*

In the parsing-based approach, the input query is first parsed, then the information generated is used to understand the grammatical structure and dependencies in the query.

*4) Grammar-based*

At the core of the grammar-based NLIDB is a set of rules that defines the questions that can be understood and answered by the system. Using rules which may have to be written by hand and are highly domain-dependent, the system can give the users suggestions on how to complete their questions during typing. This supports users to write understandable questions.

*5) Neural machine translation-based*

Neural machine translation-based NLIDB is a recent approach with a focus on applying supervised machine learning techniques on a set of query-response pairs where the queries are the natural language inputs from the user while the responses are the output SQL or SPARQL statements. This approach is highly dependent on data availability.

Research on natural language interfaces to relational databases has spanned several decades [7]. This study focuses on graph databases which excel in traversing through the nodes in a graph data by following relationships between nodes to find relevant data [11]. Many applications of the future will be built using graph databases [12]. According to Robinson et al. [14], there are three dominant graph data models, the property graph, Resource Description Framework (RDF) triples, and hypergraphs. Furthermore, graph databases such as Neo4j and JanusGraph use a variant of the property graph model. An important difference between relational and graph databases is the query language for retrieving information. While SQL is the de facto language in relational databases, a variety of declarative query languages have recently emerged for querying graph databases. SPARQL is one such language that was adopted by many vendors for querying RDF graphs while Cypher and Gremlin are the query languages for property graphs [16].

According to Park and Lim [13], a keyword-based search on a graph database usually returns a set of connected structures that represent how the data containing query keywords are interconnected in the database. The authors propose and evaluate a new ranked keyword search method for graph databases by adopting a tree-based approach in their study for efficient query processing over a large volume of graph data. They also observe that top-k answer trees based on their proposed structure and relevance measures can satisfy users' information needs better than conventional answer structures.

Oro and Ruffolo [6] designed a modular system capable of translating natural language questions into different formal queries such as SPARQL and Cypher to exploit various knowledge bases and databases. Given a specific domain, queries submitted by users contain concepts that can be categorized into ontological classes and relations.

Zhu et al. [5] propose and evaluate a natural language interface to graph-based bibliographic information retrieval. The interface can parse and interpret natural language queries by recognizing bibliographic named entities and dependency relations among the entities. The authors reported that the system can correctly answer 39 out of 40 annotated queries with varying lengths and complexities. These interfaces were fundamental to our study. Next, we describe the design of our proposed cloud-based customer query interface.

## III. SYSTEM DESIGN

### A. Design Decisions

We aimed to address the NLIDB challenges depicted in Fig. 1 and develop a proof of concept to assess the feasibility of using a multi-layered architecture with a graph database to serve queries involving multiple different distributed databases. Although in this paper we illustrate a simple use case solution involving only one relational database and a few simple queries, our architecture is designed to address multi-DB backends and complex queries which we will demonstrate in our future work. We provide a flexible chat interface to enter a natural language query and transform it into an SQL query that is executed on a backend MySQL restaurant database. Following the guidelines from Perkins et al. [11] to choose the data management and analytics use cases, we built our multi-layered solution using Neo4j as the graph database as it is open-source, fast, typeless, schemaless, and puts no constraints on relations in the data.

### B. Workflow

Our NLIDB workflow is shown in Fig. 2. A user can type a question into the system and will be returned either a ranked list of results from the main transactional database or a response indicating that the question cannot be answered. The user input is first lemmatized for improved database element selection. Parts of speeches are tagged and semantic analysis is done for noun phrase extraction. The nouns, adjectives, and noun phrases are extracted for a mapping operation. A mapping table is used to find associations between tokens and data values, while a graph database is used to find matching schema components or attributes (columns in tables) and relations (connections) between the graph nodes. Once these mappings are done, the extracted information is inserted into predefined SQL templates to formulate and execute the SQL query.

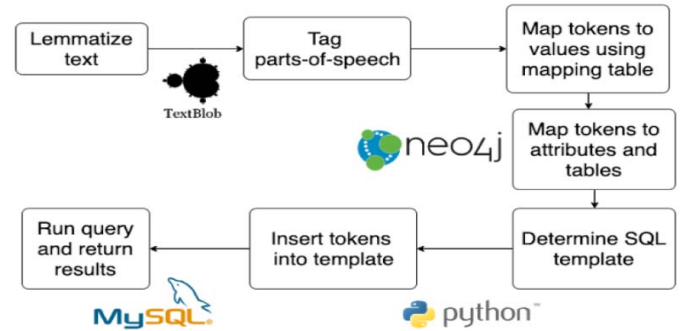

Fig.2. Data processing workflow.

## IV. SYSTEM IMPLEMENTATION

### A. Data

The data used in this project was collected from Zomato[1], a restaurant search engine, and available on Kaggle, a public data platform. The data was extracted in CSV files and inserted into MySQL. Although this system is not independent of the database, it could be adapted to other databases by refactoring the mapping table and graph database to reflect the altered schema. The schema for the SQL database is shown in Fig. 3.

### B. Implementation Details

Python 3.7 was chosen as the implementation language as its clean syntax makes it a popular choice for most data processing and analytics tasks. There are also many NLP libraries compatible with Python. TextBlob[2] was chosen as the NLP library as it is lightweight and provides various standard functions such as part-of-speech tagging and lemmatization Neo4j was chosen as the graph database to represent the schema.

---

[1] https://www.zomato.com/ncr

[2] https://textblob.readthedocs.io/en/dev/

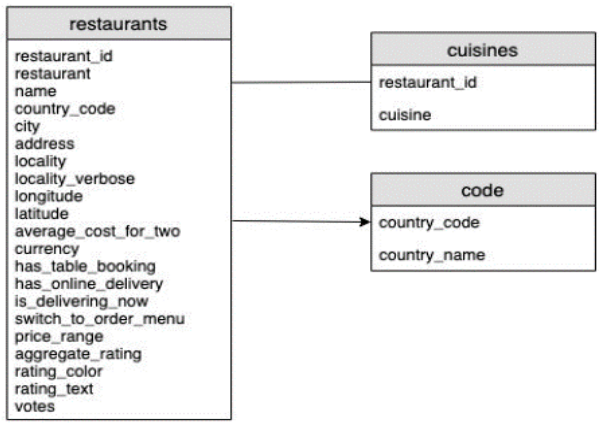

Fig.3. The database schema.

The system was implemented on a Mac OS but the source code (Python) can be easily ported to any operating systems. The PyCharm integrated development environment (IDE) was used to develop the source to help with quicker development time and fast compilation. The Python unit testing framework unittest[3] was chosen for testing as it supports test automation, sharing of setup, and shutdown code for tests.

The implementation architecture of the system is shown in Fig. 4 and consists of three layers: User Interface, Query Analysis and SQL Mapping, and Backend DBs layers. We used a simple text input in this proof of concept implementation for the user query interface which can be extended to support web-based query interface in the future. The SQL Mapping layer contains several components as described below.

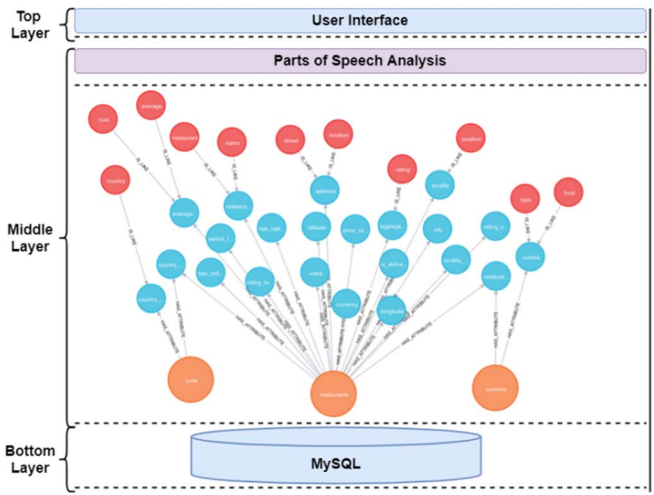

Fig. 4. The system architecture.

*1) Graph Database*

A Neo4j graph database was used to represent the schema of the backend MySQL database which can be extended to support multiple distributed and hybrid data sources. It was used to represent the tables, attributes and columns as nodes and relationships as edges. We assigned Neo4j node values as table names (e.g., code), attribute names, and synonyms according to the schema and node property values to indicate the type of schema component as table, attribute or synonym (e.g., table). Similarly, edges were also assigned values to indicate relationships and properties to indicate the types of relationships. Thus, a search through the node values based on query words (e.g., country) would lead to the matching schema component, an attribute or table or synonym, that could be used in formulating the SQL query. For example, when given the token 'country', our graph query would return all nodes having value=country and the property would indicate the node type, which would be processed further and handled based on the node type. Synonyms helped find similar terms as the query words which can be linked to an attribute or value.

---

[3] https://docs.python.org/3/library/unittest.html

Therefore, the graph in parts forms a word ontology to help map query text to SQL query which can be easily partitioned if necessary, for scalability based on the property values. Fig. 5 shows the Neo4j data model.

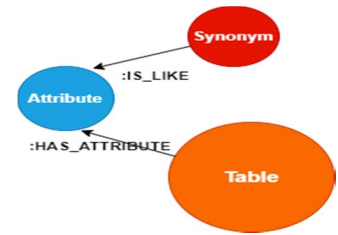

Fig.5. The Neo4j data model.

*2) Query Analysis and Mapping*

The part of the query text analysis phase consists of tokenization, mapping, and mapping table.

*a) Tokenization*

The first task of tokenization is to lemmatize the words in the given input text. Lemmatization is the process of removing inflectional endings from words and returning its base form. This transforms words such as 'restaurants' into 'restaurant' and 'deliveries' into 'delivery', making it easier for the database to correctly distinguish a concept or topic. After lemmatization, we performed part-of-speech (POS) tagging using the TextBlob NLP library to extract adjectives and nouns from the query text. Subsequently, noun phrase extraction is performed to capture multi-token semantics. To extract noun phrases from a cohesive text, a process called chunking is used to compose semantic phrases of multi-token sequences from the original text. If noun phrases are neglected, the system would not recognize words such as 'dim sum' as a cohesive entity. From the POS tagging, nouns, adjectives, and noun phrases are extracted and used in the mapping phase as these elements are most commonly used to describe database elements.

*b) Mapping*

The role of mapping is to attempt to map each token to a database element. Each token can have a set of possible corresponding elements: relation, attribute, or value. First, a mapping table is used to find if the token corresponds to a value in the database. If the value is found, then the attribute and subsequent relation will be known. If the token does not correspond to a value, it is checked to be either a relation or attribute by querying the graph database. The mapping steps for the tokens "restaurant" and "italian" is shown in Fig. 6.

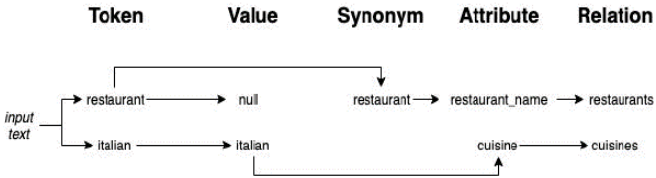

Fig. 6. The mapping steps for the tokens "restaurant" and "italian".

*c) Mapping Table*

The mapping table was designed to recognize a small number of unique values in the database from columns that would be queried often. In the future, we plan to apply machine learning algorithms to populate and update this table. We built the table using Python dictionary type which is in the form:

mapping_table [x] = y

where x is a unique value and y is the column name (attribute) that it corresponds to. Python dictionary was chosen because it has O(1) access time since the keys are accessed through a hashing function. The current mapping table contains all the unique values from the columns: cuisine, city, country_name, rating_text, currency. The mapping table in this study is relatively small (12KB) and fast to query, but with a larger database, this may become a limitation on the system resources.

*3) MySQL Database*

The SQL queries generated were restricted to the form:

SELECT {attributes} FROM {table} [, {table}] (WHERE {attribute=value} [and {attribute=value}])     …     (1)

where elements in curly braces occur once, elements in round brackets may occur once, and elements in square brackets may occur zero or more times. The mapped tokens were compiled into three lists: tables, attributes, and attribute-value pairs as follows.
1. All tables in any mapped token will be in {tables}.
2. Attributes that are not a part of an attribute-value pair will be in {attributes}.
3. All tokens which have been mapped to a table, attribute, and value will be in {attribute=value}.

We defined template strings with placeholders as shown in Eq. 1. Data from the 3 lists were used to replace the placeholders to formulate SQL queries. We used the DISTINCT keyword in the template for clarity.

As the last step, the system executed the generated SQL query on the database. Fig. 7 shows the workflow to process the natural language query "What are the restaurants and cities in India that serve fast food" and translate it into an SQL query. The text in italics describes the operation carried out at each step.

*4) Results and Validation*

The purpose of this study was to devise an algorithm to convert a natural language query into an SQL query to be executed on backend databases. Currently, the most reliable method of creating SQL queries is manual query generation by database experts. Two experiments were performed to validate our approach. The first experiment involved running multiple English queries and verifying the outputs against the human-generated SQL queries.

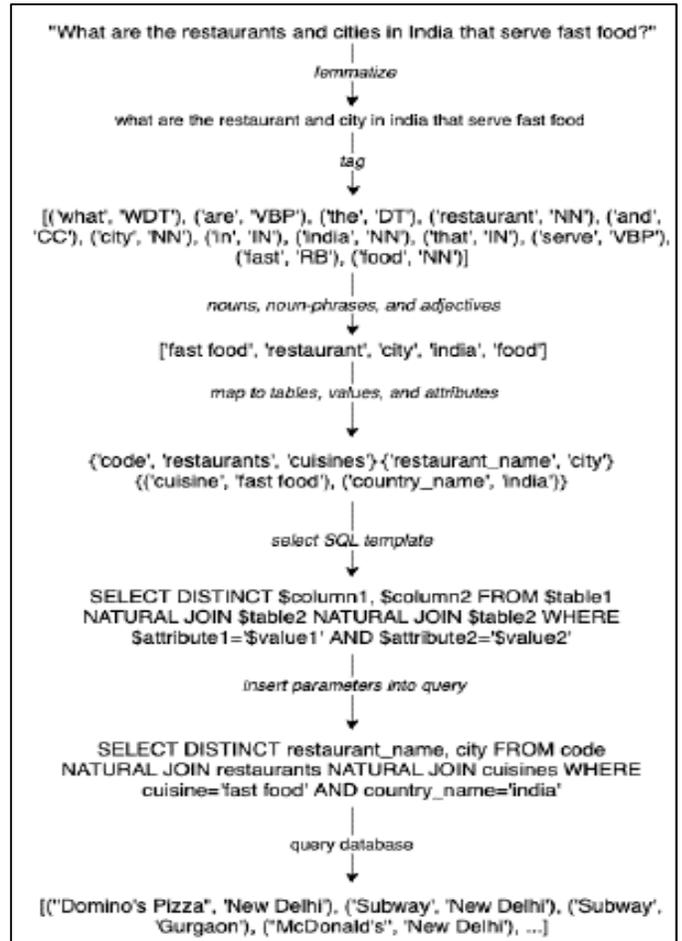

Fig. 7. A full breakdown of natural language query to SQL query.

The system was tested on numerous queries and the results of three queries are shown in TABLE I grouped by questions (Q#), human and system generated queries. Since the same query can sometimes be formulated differently, our approach was validated both quantitatively based on the accuracy in 1) retrieving the desired information, 2) extracting the correct relations, attributes, and values given the natural language query, and qualitatively based on 3) the optimality of the formulated query. Test cases were defined to validate the functionality at different phases. The translation results for ten questions yielded a 90% accuracy rate.

*5) Discussion*

The qualitative analysis for the simple correct queries proved that our approach, in comparison with previous approaches described in the literature survey section, was near-optimal. As shown in Q3, an additional column is included in the query, however, it is contextually relevant and generated the correct result. The reason behind this is that a synonym node existed in the graph database which related the word 'rating' to 'aggregate_rating', and thereby caused the selection of this column in the SQL query. Some queries did not produce correct results such as Q2, where the system generated query failed to recognize and map the adjective 'chinese' into a WHERE clause. Other queries such as Q4: "which chinese restaurants are in mumbai" also failed for the same reason, which has a similar meaning as Q2 and should produce the same SQL query and result. In Q2, the system recognized and tagged the word 'chinese' as a past participle verb, whereas for Q4 above, it was tagged as an adjective. As the system uses adjectives, nouns, and noun phrases to map to database elements, Q2 and Q4 resulted in wrong/incomplete queries. This study was exploratory to learn the challenges and develop a prototype architecture for NLI to database systems. It revealed the following key challenges which we plan to address in the future work: a) ambiguity in mapping natural language words to database schema i.e., table and column names, b) composing complex queries with multiple joins, parts and nested queries, c) distinguishing between item names and values to compose queries, and d) resolving parts of speech and error in NL query. Some of the options we would like to consider for our future work are to use an interactive NLI to resolve ambiguity, missing value and noise in query, apply machine learning methods to identify frequent queries and relationships among query items to create a rich metadata table, and extend the graph database and the architecture to support queries over hybrid distributed databases.

TABLE I: Questions (Q) and generated SQL queries

| Symbol | SQL Queries |
|---|---|
| Q1 Human | what are the italian restaurants? |
| | SELECT DISTINCT restaurant_name FROM restaurants NATURAL JOIN cuisines WHERE cuisine='italian' |
| System | SELECT DISTINCT restaurant_name FROM restaurants NATURAL JOIN cuisines WHERE cuisine='italian' |
| Q2 Human | what restaurants in mumbai have chinese food? |
| | SELECT DISTINCT restaurant_name FROM restaurants NATURAL JOIN cuisines WHERE city='mumbai' and cuisine='chinese' |
| System | SELECT DISTINCT cuisine, restaurant_name FROM cuisines NATURAL JOIN restaurants WHERE city='mumbai' |
| Q3 Human | which restaurants have an excellent rating? |
| | SELECT DISTINCT restaurant_name FROM restaurants WHERE rating_text='excellent' |
| System | SELECT DISTINCT aggregate_rating, restaurant_name FROM restaurants WHERE rating_text='excellent' |

## V. CONCLUSION

This paper reports a feasibility study on designing an NLIDB system for translating natural language queries to SQL. We define a graph model based on the schema of the backend relational database and synonymous terms, which is searched using query terms to find matching schema elements. Values in the query are searched for in a metadata table to recognize relevant schema elements. These search results are used to formulate the SQL query using predefined templates through a three-level system architecture. The test results were promising although much work is needed to support more complex queries and distributed database backends.

The future work will focus on exploring machine learning algorithms to define the metadata table, replace synonyms with existing ontologies, define complex SQL templates, for example, to support the aggregate function and nested queries such as 'how many restaurants in Canada has Mexican cuisine', and allow hybrid distributed storage systems. Recent deep learning models have shown great success in identifying phrases in natural language which can be used to process the query text to correctly identify the parts of speech and correct ill-formed queries. Although we implemented a very simple proof of concept prototype in this study, we validated the feasibility of using the graph database in a multi-tiered architecture to implement an NLIDB system that can support multiple backend databases. The architecture can be further leveraged using machine learning techniques to learn query patterns, frequent queries and cache the responses to reduce response time and improve the overall system performance. Artificial intelligence conversation techniques can be used to enable interactive query processing which can also help disambiguate the query objectives and enable prediction of the next query for efficient query processing.